\documentclass[aps,prd,reprint,nofootinbib,groupedaddress,preprintnumbers,longbibliography]{revtex4-1}

\usepackage{amsmath,amssymb,mathtools,bm}
\usepackage{graphicx, color}
\usepackage[dvipsnames]{xcolor}
\usepackage{float}
\usepackage[multiple]{footmisc}

\definecolor{c1}{HTML}{802410} 
\definecolor{c2}{HTML}{003262} 
\definecolor{c3}{HTML}{00A598}
\definecolor{c4}{rgb}{1., 0.498039, 0.054902}

\newcommand\Star[3][]{%
\path[#1] (0  :#3) -- ( 36:#2) 
       -- (72 :#3) -- (108:#2)
       -- (144:#3) -- (180:#2)
       -- (216:#3) -- (252:#2)
       -- (288:#3) -- (324:#2)--cycle;}

\usepackage{tikz}
\usepackage{tkz-euclide}
\usetikzlibrary{decorations.pathmorphing}	
\tikzset{
    v/.style={decorate, decoration={snake, segment length=3mm, amplitude=0.75mm}, draw},
    f/.style={draw,decoration={markings,mark=at position #1 with {\arrow[very thick]{latex}}},postaction={decorate},node contents=#1},
    f/.default=.6,
    fb/.style={draw,decoration={markings,mark=at position #1 with {\arrowreversed[very thick]{latex}}},postaction={decorate},node contents=#1},
    fb/.default=.4,
    fnar/.style={draw},
    g/.style={decorate, draw,  decoration={coil,amplitude=3pt, segment length=3.5pt}},
    s/.style={dashed,draw, postaction={decorate},
        decoration={markings,mark=at position .55 with {\arrow[very thick]{latex}}}},
    sb/.style={dashed,draw, postaction={decorate},
        decoration={markings,mark=at position .55 with {\arrowreversed[draw=black,very thick]{latex}}}},
    snar/.style={dashed,draw,line width =1.25pt},
}
\usetikzlibrary{shapes}	
\tikzset{every picture/.style={line width=1}}
\usepackage{mathrsfs}
\newcounter{qnumber}

\usepackage{tensor}

\usepackage{orcidlink}

\usepackage{hyperref} 
\hypersetup{
    colorlinks=true,      
    linkcolor=blue,        
    citecolor=blue,        
    filecolor=magenta,     
    urlcolor=blue          
}

\begin{document}

\title{Astrometric Detection of Ultralight Dark Matter}

\author{Jeff A. Dror \orcidlink{0000-0003-0110-6184}}
\email{jeffdror@ufl.edu}
\affiliation{Institute for Fundamental Theory, Physics Department, University of Florida, Gainesville, FL 32611, USA}

\author{Sarunas Verner \orcidlink{0000-0003-4870-0826}}
\email{verner.s@ufl.edu}
\affiliation{Institute for Fundamental Theory, Physics Department, University of Florida, Gainesville, FL 32611, USA}
\vspace{0.5cm}

\date{\today}

\begin{abstract}
Ultralight dark matter induces time-dependent perturbations in the spacetime metric, enabling its {\em gravitational direct detection}. In this work, we propose using astrometry to detect dark matter. After reviewing the calculation of the metric in the presence of scalar dark matter, we study the influence of the perturbations on the apparent motion of astrophysical bodies. We apply our results to angular position measurements of quasars, whose vast distances from Earth present an opportunity to discover sub-component dark matter with a mass as low as $10^{-33} \, \mathrm{eV} $. We explore the prospects of very long baseline interferometry and optical astrometric survey measurements for detecting ultralight relics, finding that for the smallest masses, current astrometric surveys can detect dark matter moving locally with a velocity of $10^{-3}$ with energy density as low as $10 ^{ - 4} ~{\rm GeV} / {\rm cm} ^3 $.
\end{abstract}

\maketitle

\vspace{0.2cm}
\noindent \textbf{Introduction.} 
To date, all successful attempts to infer the presence of dark matter have relied solely on its gravitational interactions with the visible sector. What if dark matter only interacts gravitationally? It was recently demonstrated that, in this case, it is still possible to {\em directly detect} dark matter through time-dependent perturbations to the spacetime metric generated by its coherent oscillations~\cite{Khmelnitsky:2013lxt}. These perturbations cause a gravitational redshift that can be measured by analyzing pulsar timing array (PTA) data, provided dark matter has a de Broglie wavelength shorter than the distance to typical millisecond pulsars, corresponding to a mass $ m \gtrsim 10 ^{ - 24} ~{\rm eV} $. Experimental searches have since been conducted by the Parkes PTA~\cite{Porayko:2018sfa}, the European PTA~\cite{EPTA:2023xiy}, and NANOGrav~\cite{NANOGrav:2023hvm}. Concurrently, there is an ongoing large-scale theory effort to understand the detection prospects of the dark matter-induced gravitational redshift using gravitational wave instruments (see, e.g., Refs.~\cite{Porayko:2014rfa, Graham:2015ifn, Aoki:2016mtn, Aoki:2016kwl, Blas:2016ddr, DeMartino:2017qsa, Kato:2019bqz, Nomura:2019cvc, Kaplan:2022lmz, Xia:2023hov, Luu:2023rgg, Kim:2023pkx, Hwang:2023odi, Kim:2023kyy, Brax:2024yqh, Yu:2024enm}).

In this \textit{Letter}, we propose a new method for the gravitational direct detection of dark matter: measuring the apparent angular positions of astrophysical bodies, a technique known as astrometry. Astrometry has seen significant advancements with the maturation of very long baseline interferometry (VLBI), which achieves positional uncertainties as low as 1~$ \mu $as for a small number of sources, and the advent of space-based optical telescopes, which, despite having higher positional uncertainties, provide recurrent measurements of billions of sources~\cite{Reid:2013gva, 2017isra.book.....T}. 

For an observer moving at a non-relativistic instantaneous velocity, $ {\mathbf{v}} _{\rm o} $, with a constant acceleration, $ {\bf a} _{\rm o}  $, the angular position of an astrophysical source moving at $ {\mathbf{v}} _{\rm s} $ at a far distance $ {\mathbf{x}} _s $ exhibits known kinematic corrections. The apparent proper motion in an angular direction $ \hat{ \theta }$ is,
\begin{equation} 
\omega _\theta = \left[ \dot{ {\mathbf{v}}} _{\rm o} - \frac{{\mathbf{v}} _{\rm o} - {\mathbf{v}} _{\rm s}  }{ x _{\rm s} } + {\bf a} _{\rm o}  \right] \cdot \hat{ \theta } \,,
\label{eq:propermotion}
\end{equation}
with the terms known as classical aberration, intrinsic proper motion, and secular aberration drift, respectively. In Eq.~\eqref{eq:propermotion}, we only present the leading term for each type of correction. 

Classical aberration is sizable for every source, inducing a typical proper motion of $ {\cal O} ( 100~ {\rm as}/ {\rm year} ) $ for observers orbiting the Sun, and is routinely corrected for (see, e.g., Ref.~\cite{2003AJ....125.1580K}). For extragalactic sources (the focus of this work), the second term in Eq.~\eqref{eq:propermotion} is negligible.\footnote{For a source at a cosmological distance, the expression for intrinsic proper motion is identical with relative velocity interpreted as the peculiar velocity and $ x _s $ as the comoving distance.} For the Milky Way potential, the acceleration-induced proper motion is measured to be $ 4.80 \pm 0.08~\mu {\rm as}/{\rm year} $~\cite{Titov:2010zn,2021A&A...649A...9G}, displaying a dipolar pattern across the sky. 

Ultralight dark matter also induces an apparent angular deflection independent of the source distance.\footnote{Gravitational waves with very low frequencies may also induce substantial astrometric deflections independent of source distance, as was considered in Refs.~\cite{Braginsky:1989pv, Kaiser:1996wk, Pyne:1995iy, Book:2010pf, Darling:2018hmc,Mihaylov:2018uqm,  Garcia-Bellido:2021zgu, Fedderke:2022kxq, Liang:2023pbj, Caliskan:2023cqm}.  An astrometric search for ultralight vector bosons coupled to a B-L current was proposed using Gaia data~\cite{Guo:2019qgs}.} We calculate the effect in this work, finding a characteristic signal of secular proper motion analogous to the secular aberration drift. Importantly, the size of the effect depends on the hierarchy between the source distance and the inverse dark matter mass so that, for a large range of masses, galactic and extragalactic sources exhibit a different deflection. This provides a method to distinguish the signal from other backgrounds. 

For a fixed density, the dark matter-induced angular deflection is most pronounced at the lowest dark matter masses. The cosmological distances of quasars enable the detection of dark matter with masses as low as the current Hubble constant.\footnote{While such masses conflict with the conventional lower bound of approximately $ 10^{-22} - 10^{-20} ~\rm{eV}$~\cite{Kobayashi:2017jcf, Irsic:2017yje, Nori:2018pka, Leong:2018opi, Schutz:2020jox, DES:2020fxi, Rogers:2020ltq,Dalal:2022rmp}, these cosmic relics might exist as a sub-component of the observed dark matter density.} As such, we focus on the detection of dark matter using quasars, though most of our expressions can be applied to astrometry of any point source.

Our findings pave the way for astrometry to revolutionize the prospects of gravitational direct detection of dark matter. As we show, this method is sensitive to minuscule densities and is a powerful probe of the existence of ultralight particles.  

\vspace{0.2cm}
\noindent \textbf{Ultralight Scalar Dark Matter.} 
Ultralight scalar dark matter can be modeled as a classical field, $ a ( t ,{\mathbf{x}} ) $, oscillating primarily at a frequency $\Omega$ with a slowly varying phase, $\alpha(t, \mathbf{x})$, 
\begin{equation}
    \label{eq:ansatzfull}
    a (t, \mathbf{x})  = a_0(\mathbf{x}) \cos \left(\Omega t + \alpha(t, \mathbf{x}) \right) \, .
\end{equation}
The velocity distribution of the field is encapsulated in the properties of $ \alpha ( t, {\mathbf{x}} )  $ (see, e.g., Refs.~\cite{Foster:2020fln,Dror:2022xpi} for further details). This phase can be decomposed as $ - m  {\mathbf{v}} _a    \cdot {\mathbf{x}} + \alpha _0  $, where $ {\mathbf{v}} _a $ represents the field velocity relative to the cosmic frame. The offset $ \alpha _0 $ and the direction of $ {\mathbf{v}} _a $ are randomly sampled every coherence time and coherence length of the field (both quantities determined by the dark matter velocity distribution). The momentum of the field, $ {\mathbf{k}} \equiv \nabla \alpha = - m  {\mathbf{v}} _a  + \nabla \alpha _0 $, is related to the frequency through the Klein-Gordon equation, $\left(\Box - m^2 \right)a(t, \mathbf{x}) = 0$ (in the non-relativistic limit, $ \Omega =  m  + k ^2 / 2 m $). The conditions under which the field can be accurately approximated by a fluctuating plane wave form are $ k\gg \nabla a _0 / a _0   $ and $ \Omega  \gg \partial _t  \alpha  $, and we assume these conditions throughout. 

Ultralight dark matter generates time-dependent perturbations in the stress-energy tensor, resulting in oscillating perturbations to the spacetime metric through Einstein's equations~\cite{Khmelnitsky:2013lxt}. The stress-energy tensor of a scalar field in an otherwise flat spacetime is given by:
\begin{equation} 
  \tensor{ T }{ _\mu  _\nu }= \partial _\mu  a \partial _\nu a - \tensor{ \eta  }{ _\mu  _\nu } \left( \frac{1}{2} \eta  ^{ \sigma \rho  }\partial _\sigma  a \partial _\rho  a + \frac{1}{2} m ^2 a ^2  \right)\,,\label{eq:T}
\end{equation} 
where $ \eta \equiv {\rm diag} ( - 1 , 1, 1, 1 ) $. Inserting the form of $ a  ( t ,{\mathbf{x}} ) $ from Eq.~\eqref{eq:ansatzfull} into this expression yields oscillating energy density $\rho \equiv T_{00}$ and pressure $P \equiv \frac{1}{3}T_{ii}$ terms at a frequency of $ 2 \Omega $:
\begin{align}
    \rho & =   \frac{1}{2} a_0^2 \left[m^2 + k^2 - k^2 \cos \left(2 ( \Omega t + \alpha )  \right) \right] \,,     \label{eq:enden2a}
\\
    P & =  \frac{1}{2}a_0^2 \left[\frac{1}{3}k^2 - \left(m^2 + \frac{1}{3}k^2 \right) \cos \left(2 ( \Omega t + \alpha )  \right) \right] \,.
     \label{eq:pressure2a}
\end{align}
In deriving these expressions, we neglected $\nabla a_0$ and $\partial_t \alpha$ terms (see discussion above) and terms of $ {\cal O} ( k ^4 ) $. The results agree with Ref.~\cite{Khmelnitsky:2013lxt} at $ {\cal O} ( k ^0 ) $. Since the Einstein equations are second-order differential equations, it is useful to retain the (naively) higher-order terms. While our interest is in experiments that resolve the time oscillation, we note that, upon time-averaging, the pressure vanishes and the energy density takes the familiar form, $\rho \simeq \frac{1}{2}  m^2   a_0^2$. 

The oscillatory nature of the stress-energy components generates time-dependent scalar perturbations in the spacetime metric. The scalar-vector-tensor decomposition theorem (see, e.g., Ref.~\cite{Maggiore:2018sht}) ensures that the scalar perturbations generate scalar metric perturbations. We note that scalar-vector-tensor perturbations do not mix at leading order. Since we only have scalar stress-energy perturbations, the influence of dark matter is to induce scalar metric perturbations. 

In Newtonian gauge, the metric is characterized by two scalar functions, $\phi$ and $ \psi $, such that the line element is given by,
\begin{equation}
\label{eq:pertnewt}
d s^2 = -(1+2 \psi (t, {\bf{x}})) d t^2 + \left(1 - 2 \phi  (t, {\bf{x}}) \right) \delta_{ij} d x^i d x^j \, .
\end{equation}
These gravitational potentials decompose into time-independent and time-dependent contributions with frequency $ 2 \Omega $~\cite{Khmelnitsky:2013lxt}:
\begin{align} 
\label{eq:osc1}
  \psi  & = \psi _0+\psi _c \cos \left(2 ( \Omega t + \alpha )  \right) + \psi _s \sin \left(2 ( \Omega t + \alpha )  \right) ,  \\
\phi & =  \phi_0+\phi_c \cos \left(2 ( \Omega t + \alpha )  \right) + \phi_s \sin \left(2 ( \Omega t + \alpha )  \right) \, ,
\end{align} 
where the coefficients $ \psi _0 ,\psi _c , \psi _s , \phi _0 ,\phi _c , \phi _s $ depend only on $ {\mathbf{x}} $. Combining Einstein equations in Newtonian gauge,\footnote{The Christoffel symbols are,
\begin{center} \begin{tikzpicture} 
\node at (0,0) {$ \Gamma _{ 0 0 } ^0   =  \dot{ \psi  } \,,$}; 
\node at(2.5,0) {$ \Gamma _{ i0} ^0  = \Gamma _{ 0 i } ^0 = \Gamma _{ 0 0 } ^i = \partial _i \psi   \,,$};
\node at(6,0) {$ \Gamma _{ ij } ^0   = \Gamma _{ 0 j} ^i  =\Gamma ^i _{ j 0} =    - \dot{ \phi  }  \delta _{ij}\,,$};
\node at(3,-0.5) {$ \Gamma _{ij} ^k  =  \partial _k \phi \delta _{ij} - \partial _i \phi  \delta _{ jk} - \partial _j \phi  \delta _{ ik}\,.$};
\end{tikzpicture}
\end{center}
}
\begin{equation}
    \nabla^2 \phi  =  4 \pi G \rho   \, , \quad \ddot{\phi} + \frac{1}{3}\nabla^2 \left(\psi - \phi \right)  =   4 \pi G P  \, ,
\end{equation}
with Eqs.~(\ref{eq:enden2a}) and (\ref{eq:pressure2a}), and separating the time-dependent and time-independent contributions, we find that the gravitational potentials are:
\begin{align}
\psi_c(\mathbf{x})  =  - \phi _c ( {\mathbf{x}} )  = -\frac{1}{2}\pi G a_0^2 (\mathbf{x}) \,,  \phi_s = \psi_s = 0\,,
\label{eq:phipotfullX}
\end{align}
where $ G $ denotes the gravitational constant. $\phi_0(\mathbf{x})$ and $\psi_0(\mathbf{x})$ describe the static gravitational potential and are indistinguishable from those of cold dark matter, which we neglect in our analysis. In the following analysis, we focus on dark matter by replacing $ \psi $ with $ - \phi  $.

\vspace{0.2cm}
\noindent \textbf{Astrometric Deflection of Dark Matter.} 
The dark matter metric perturbations cause fluctuations in the apparent positions of distant astrophysical sources. We now derive this effect working to first order in the size of the perturbation, assuming they are superimposed on a flat Minkowski background. Although our results can be readily generalized to a Friedmann-Robertson-Walker (FRW) metric via a conformal transformation~\cite{Book:2010pf}, the results remain unchanged in the limit when $ \Omega $ is large compared to the inverse source distance. For quasars (typically at cosmological distances), this holds for all masses well above the Hubble scale today. 

In the absence of metric perturbations, light emitted from a source with frequency $ \omega _0 $ traveling in the $ - \hat{\bf n} $ direction has the following position and momentum four-vectors:
    \begin{align}
   x^{\mu}(\lambda)  & = \left(\omega_0 \lambda + t _{ {\rm o}}, -\omega_0 \lambda \hat{\bf{n}} \right) \, , \\
   p^{\mu}(\lambda)  & = \omega_0 \left(1, -\hat{\bf{n}} \right) \, .
   \end{align}
Here, $\lambda$ is the affine parameter, with a value of $ \lambda _{ {\rm s}} \equiv   - \left| {\mathbf{x}} _{ {\rm s}} \right| / \omega _0 $ corresponding to photon emission and $  0 $ at photon arrival, while $t_{ {\rm o}}$ represents the time of photon observation. We choose the observer to be at rest at the origin upon observation, and the source at rest (our results are insensitive to this final assumption), with position four-vectors $ x^{\mu}_{\rm o}(t) = (t,  {\mathbf{0}}   ) $, $ x_{\rm s}^{\mu}(t) = (t, {\bf x}_{\rm s})$.  Our goal is to solve for the angular deflection. Since both the observer and source are at rest, we have $ u_{\rm o}^{\mu} = u_{\rm s}^{\mu} =  ( 1 , {\mathbf{0}})$.\footnote{The location of the observer and the vector pointing toward the observer are formally {\em time-varying} quantities. Corrections to the final result due to this variation are suppressed by the ratio of the displacement of the object and the distance to the source and are typically quite small. We neglect these throughout.}

We introduce a coordinate system for the observer in the absence of metric perturbations, $ ( e ^\mu _{ \hat{ 0}}, e ^\mu _{ \hat{ 1}},e ^\mu _{ \hat{ 2}},e ^\mu _{ \hat{ 3}} ) $. We set the basis vector equal to the observer's four-velocity, $e_{\hat 0}^{\mu} = (1, \mathbf{0})$ since the observer is at rest. Requiring the basis be orthogonal ($\eta_{\mu \nu} e_{\hat{\alpha }}^\mu e_{\hat{\beta }}^\nu  = \eta_{\hat{\alpha } \hat{\beta }}$) implies $e ^i _{ \hat{ j}} = \delta ^i _{\hat{ j}} $ and $ e _{ \hat{ a }} ^0  =  0$. Here and throughout, the hatted indices denote the observer frame and the unhatted indices denote the coordinate frame. 

The apparent four-momentum of the source in the observer frame is given by $  p _{ \hat{ \alpha  }}  =  \eta _{ \mu \nu }  p ^\mu  e ^\nu _{ \hat{ \alpha }}  $, which gives an observed photon frequency and 3-momentum, 
\begin{equation} 
p _{ \hat{ 0}} = - \omega _0  \,, \, p _{ \hat{a}} =  - \omega _0 \hat{\bf n}  \,.
\end{equation} 

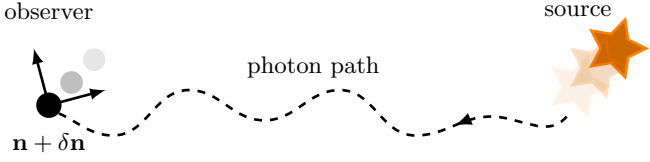
\begin{figure}
\begin{center} \begin{tikzpicture} 
\coordinate(observer) at (0,0);
\coordinate(source) at ($ (observer)+(7,0) $);
\coordinate (O) at ($ (observer)+(-0.,0.) $);
\draw[fill=black] (observer)  circle (0.15);
\draw[fill=black,opacity=0.25,draw=none] ($ (observer)+(0.3,0.3) $)  circle (0.15);
\draw[fill=black,opacity=0.1,draw=none] ($ (observer)+(0.6,0.6) $)  circle (0.15);
\begin{scope} [xshift=7.5cm,yshift=0.75cm]
\Star[fill=orange!80!black,draw=orange]{.2}{.4} 
\end{scope}
\begin{scope} [xshift=7.25cm,yshift=0.5cm]
\Star[fill=orange!80!black,draw=orange,opacity=0.25]{.2}{.4} 
\end{scope}
\begin{scope} [xshift=7cm,yshift=0.25cm]
\Star[fill=orange!80!black,draw=orange,opacity=0.1]{.2}{.4} 
\end{scope}

\draw[f=0.2,dashed] ($ (source)+(-0.15,-0.15) $) to [out=-130,in=0] ++ (-1,0) to [out=180,in=0] ++ (-1,-.25) to [out=180,in=0] ++ (-1,.6) to [out=180,in=0] ++ (-1,-0.4)to [out=180,in=0] ++ (-1,0.4)to [out=180,in=0] ++ (-1,-0.6) to [out=180,in=0] ++ (-1,0.35) ;
\draw[-latex] (observer) --++(0.75,0.2);
\draw[-latex] (observer) --++(-0.2,0.75);
\node at($ (observer)+(0,-0.5) $) {$ {\mathbf{n}} + \delta {\mathbf{n}} $};
\node at ($ (observer)!0.5!(source)+(0,.5)  $) {photon path};
\node at ($ (observer)+(0,1.25) $) {observer};
\node at ($ (source)+(0,1.25) $) {source};
  \end{tikzpicture}
\end{center}
    \caption{Depiction of the influence of dark matter on astrometry: In a general gauge, dark matter affects the photon path, the locations of the source and observer, and the observer's reference frame.}
    \label{fig:gwgeometry}
\end{figure}

At first order in the perturbations, there are three {\em gauge-dependent} effects: the photon worldline and momentum perturbations ($ \delta x ^\mu ( \lambda )  $ and $ \delta p ^\mu ( \lambda )  $), the source and the observer worldline perturbations ($ \delta x _{\rm s} ( \lambda   ) $ and $ \delta x _{ {\rm o}} ( \lambda  ) $), and the frame deformation in the local proper reference frame of the observer, $ \delta e _{ \hat{ \alpha }} ^\mu (\lambda) $. The trajectory of the photon in the presence of perturbations is depicted in Fig.~\ref{fig:gwgeometry}. To find the deflection of the source, we calculate its four-momentum in the observer frame,
\begin{equation} 
 p _{ \hat{ \alpha  }} + \delta p _{\hat{\alpha } } =  ( \eta _{ \mu \nu } \hspace{-0.025cm}+\hspace{-0.025cm} h _{ \mu \nu } ( 0 )  ) ( p ^\mu \hspace{-0.025cm}+\hspace{-0.025cm} \delta p ^\mu ( 0 ) )  ( e ^\nu _{ \hat{ \alpha}} + \delta e ^\nu _{ \hat{ \alpha }} (0) )  \,.
\label{eq:observedp}
\end{equation} 
The apparent direction of the source, $\hat{n}_{\hat i} + \delta \hat{n}_{\hat i} = (p_{\hat{i}} + \delta p_{\hat i} ( 0 ) )/(p_{\hat{0}} + \delta p_{\hat 0} ( 0 ) )$, is gauge-independent. Its calculation requires expressions for $ \delta p ^\mu ( 0 ) $ and $ \delta e ^{\mu } _{ \hat{ \alpha }} (0)$.

The first-order correction to the photon four-momentum can then be computed using the geodesic equation,
\begin{equation}
    \label{eq:geopertmom}
    \frac{d \delta p^{\mu}}{d \lambda} \; = \; - \Gamma^{\mu}_{\nu \rho} (\lambda) p^{\nu} p^{\rho} \, ,
\end{equation}
where the Christoffel symbols are evaluated along the trajectory of the photon at zeroth order. Since the time and space component differential equations become analogous in the limit where $ \psi = - \phi $, the computation is significantly simplified. Below, we present the key results, with the full derivation provided in the Supplemental Material.

From the photon geodesic equation~(\ref{eq:geopertmom}), we find that the observed photon four-momentum perturbation is given by
\begin{align}
     \delta p^0(0) &=  \omega_0 \left[2 \phi(0) - \phi(\lambda_s) \right] + \omega_0^2 \int_{\lambda_{\rm ref}}^{\lambda_s} \hat{\mathbf{n}} \cdot \nabla \phi(t(\lambda'), \mathbf{x}_s) d \lambda'  , \\
     \delta {\mathbf{p}} (0)  & =   \omega_0 \left[\phi(\lambda_{ {\rm s}})  -2\phi(0) \right] \hat{\bf n}  - \omega_0^2 \hat{\bf n} \int^{\lambda_s   }_{\lambda _{\rm ref}} \hspace{-0.2cm} \hat{\bf n} \cdot  \nabla \phi  (t ( \lambda '), {\mathbf{x}} _{\rm s}  )d \lambda ' ~\notag \\
     \label{eq:deltap0}
 &-  \frac{1}{ \lambda_{\rm s}}\hat{\bf n} \times \left[ \hat{\bf n} \times \left[ \delta {\mathbf{x}} _{\rm s} ( \lambda _{\rm s} ) - \delta {\mathbf{x}}  _{\rm o} (0) \right] \right] \, ,
\end{align}    
where the lower bound on the integral is taken at an unphysical reference point, $\lambda _{\rm ref}$, and we have fixed the integration constants by requiring the photon left the source at $\lambda = \lambda _{\rm s}$ and reached the observer at $\lambda=0$.

We determine the perturbation to the spatial basis vectors using the parallel transport equation at first order in the perturbation size,
\begin{equation}
\label{eq:geodesicfore}
\frac{1}{ \omega _0 } \frac{ d \delta e ^\nu _{ \hat{ a}} }{ d \lambda  } = -  \Gamma ^\nu _{ 0 \alpha } e ^\alpha _{ \hat{ \alpha } } \,,
\end{equation} 
where the Christoffel symbols are evaluated along the observer worldline. Solving these equations leads to
\begin{align} 
\delta e_{\hat{a}}^i & =   \phi (t, {\mathbf{0}} ) \delta_{\hat{a}}^i \,,\\ 
\delta e ^0 _{ \hat{ a}}   & = \omega _0 \int _{ \lambda _{\rm ref}} ^\lambda   \hspace{-0.1cm} \partial _{ \hat{ a }} \phi   ( t' ( \lambda )  ,{\mathbf{0}} ) d \lambda ' \,,
\end{align} 
up to omitted integration constants which are time-independent and unobservable.

\begin{figure*}[t!]
    \centering
    \includegraphics[width=0.7\textwidth]{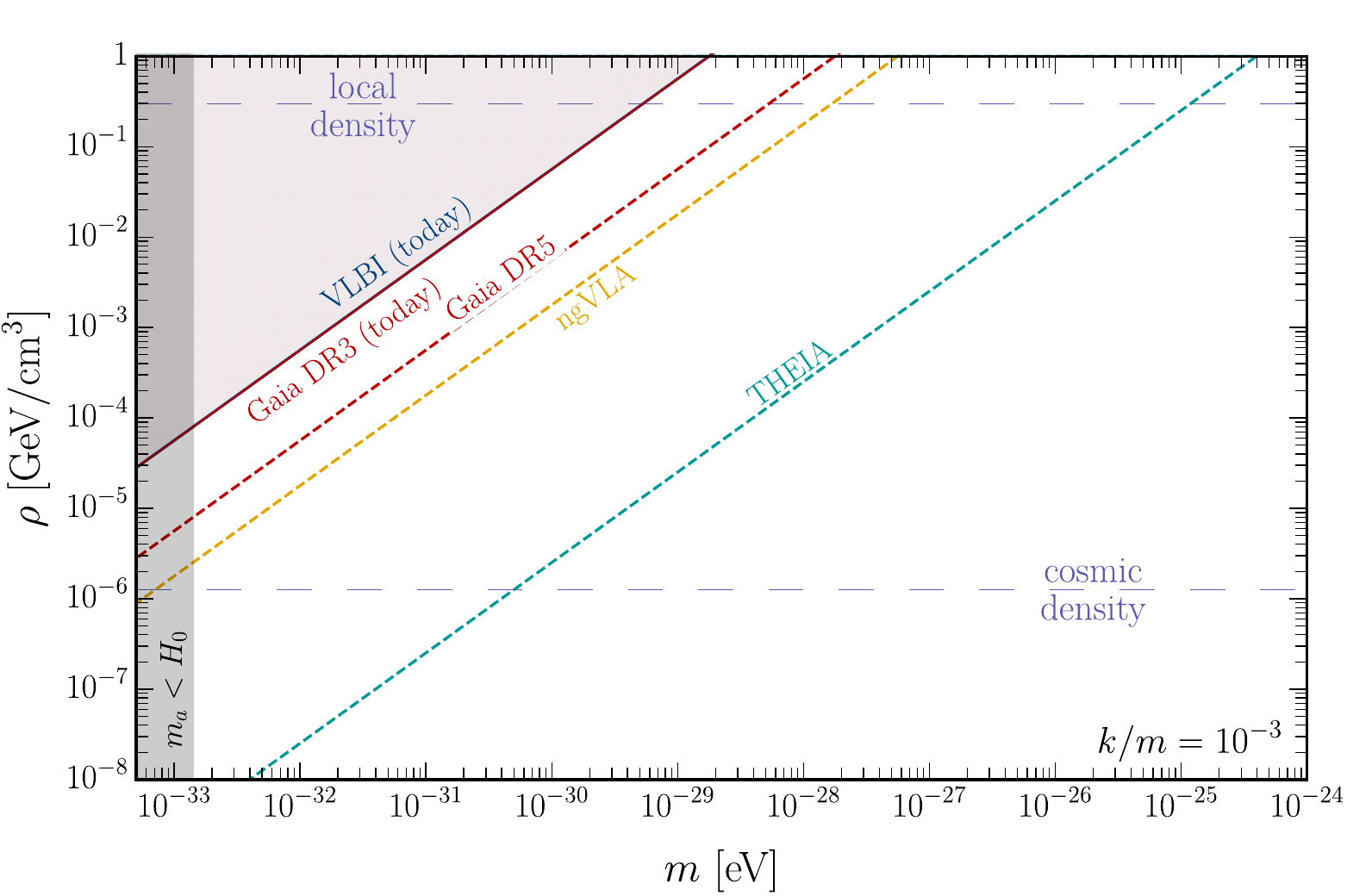}
    \caption{The sensitivity of astrometry to ultralight dark matter. Solid lines represent the potential of reanalyses of existing VLBI (solid blue) and Gaia DR3 (solid red) data to detect dark matter. Dashed lines project sensitivities for near-future data releases from Gaia DR5 (dashed red) and the next-generation Very Large Array (dashed yellow). The sensitivity of a search using THEIA, a proposed successor to Gaia, is shown in dashed teal. Details of the chosen parameters are discussed in the main text.}
    \label{fig:constraints}
\end{figure*}

We now have all the ingredients to calculate the observed photon four-momentum using Eq.~\eqref{eq:observedp}. We are interested in {\em angular} deflection, so terms proportional to $ \hat{\bf n} $ drop out. The deflection in the $ \hat{\theta}  $ direction is given by, 
\begin{align} 
\delta {\hat{\mathbf{n}}} \hspace{-0.025cm}\cdot \hspace{-0.025cm}\hat{\theta}   &\hspace{-0.05cm}= \hspace{-0.05cm} \left[ \delta e ^0 _{ \hat{a} } - \hspace{-0.05cm} \frac{\delta p ^j \hspace{-0.025cm} (0)  \delta _{ \hat{ a}} ^j   }{ \omega _0 } \right] \hat{ \theta}  ^{\hat{ a} } . \label{eq:deltan2}
\end{align} 
Our primary interest is in searching for dark matter with a mass well below any inverse experimental operation time, $ m t \ll 1 $. In this limit, not every contribution within Eq.~\eqref{eq:deltan2} is differentiable from the kinematic aberrations and intrinsic proper motion terms of Eq.~\eqref{eq:propermotion}. To extract the physical influence of dark matter, we calculate the difference in deflections exhibited by a source obeying $x _{\rm s} \ll m ^{-1} $ and one obeying $ x _{\rm s} \gg m ^{-1} $,\footnote{We always assume that $ x _{\rm s} $ is large compared to the displacement of the observer in the absence of dark matter.} 
\begin{equation}
\begin{aligned} 
\label{eq:deflectiondifference}
\Delta (  \delta  \hat{\bf n}  )  \cdot \hat{\theta} &  = \omega _0 \int _{ \lambda _{{\rm ref}} } ^0 d \lambda \nabla \phi ( t ( \lambda ) , {\mathbf{0}} ) \cdot \hat{\theta}  \,, \\ 
& = \frac{ k \phi _c }{ m } \cos ( 2 ( \Omega  t + \alpha ) ) \,,
\end{aligned} 
\end{equation}
where we have dropped a small contribution suppressed by $  m ^2 $ and assume, for simplicity, a dark matter gradient oriented along the $\hat \theta$ direction. Remarkably, this difference is independent of the specific distances to the sources.

We find that the presence of ultralight dark matter induces a secularly-varying aberration resulting in a time-independent proper motion,\footnote{Recent work~\cite{Mishra-Sharma:2020ynk} has demonstrated that astrometry can be used to probe galactic substructure through induced deflections, velocities, and accelerations. However, because the deflection in the distant source limit is only sensitive to the local dark matter density, astrometry is insensitive to dark matter density variations.}
\begin{equation}
 k \phi _c \hspace{-0.05cm}\sim     \hspace{-0.05cm} \frac{ 0.3~\mu {\rm as} }{ {\rm year}}\left( \frac{ 10 ^{ - 29}~ {\rm eV} }{ m} \right)  \hspace{-0.1cm}\left( \frac{ k /m }{ 10 ^{ - 3} }\right)    \frac{\rho}{0.3 \, \frac{ {\rm GeV} }{  {\rm cm} ^3}}   \,.
 \label{eq:omega}
\end{equation}

\vspace{0.2cm}
\noindent \textbf{Detection Prospects and Discussion.} 
To estimate the sensitivity of astrometric observations to dark matter, consider measurements of $N_q$ quasars, observed at regular time intervals $\Delta t$, for a total time $T$, with an instrumental uncertainty of $\sigma_\theta$. In the presence of dark matter, the apparent quasar locations exhibit correlated proper motions given by Eq.~\eqref{eq:omega}. We estimate the sensitivity an optimal analysis for each search can have to dark matter using the log-likelihood ratio test. The details of the sensitivity estimate are described in the Supplemental Material. 

If no signal is present in the data, we find projected exclusion limits,
\begin{equation}
k \phi_c < \sigma_\theta\sqrt{\frac{6 q_{\rm th}}{N_q}} \frac{1}{T}\sqrt{\frac{ \Delta t}{T}} \,, 
\label{eq:qth}
\end{equation}
where $q = q_{\rm th} \simeq 2.7$ corresponds to a 95\% upper limit.  The sensitivity expressions assume only instrumental noise is present in the data and hence the signal can be completely distinguished from proper motion and secular aberration. This could be done by exploiting the distance dependence of the dark matter-induced aberration; nearby sources could be used to ``calibrate" the search, and quasars could be used to search for dark matter. 

Measurements of 713 quasars were recently compiled with up to $ 1 - 10 ~\mu {\rm as} $-level precision per measurement from a combination of archival VLBI data and recent results from the Very Large Baseline Array (VLBA)~\cite{Truebenbach_2017}. The VLBA measurements represent a dramatic improvement in sensitivity, presenting a powerful opportunity to test for the existence of ultralight dark matter. To estimate the sensitivity achievable with current data, we assume $ 50 $ well-measured quasars, each observed yearly with $ 5 ~ \mu {\rm as} $ precision/measurement for ten years. 

The sensitivity of VLBI will improve dramatically over time with an increasing number of measurements. Additionally, there will be a considerable gain in sensitivity if the next generation Very Large Array (ngVLA) project is constructed~\cite{2018AAS...23134202M,2019clrp.2020...32D,Kadler:2023mjd}. Ref.~\cite{2018ASPC..517..523R} suggested that ngVLA could measure $30$ times more sources while reaching $1~\mu {\rm as} $ sensitivity. To estimate the sensitivity of ngVLA, we assume $ 1000 $ sources observed twice per year for a decade at this precision. 

The Gaia space observatory measures a much larger set of extragalactic objects but with a precision of around 200 $ \mu $as~\cite{Gaia}. Gaia's most recent DR3 data release amounts to 34 months of data with multiple measurements per extragalactic source, totaling approximately 1 million sources~\cite{2021A&A...649A...9G}. To estimate the sensitivity, we assume 10 measurements per source. For the full Gaia DR5 dataset, we assume a factor of two improvement in precision, $ 2 \times 10 ^{ 6} $ sources, a 10-year dataset, and yearly measurements for each source. Looking further ahead, we also project the capabilities of a potential successor to Gaia, THEIA~\cite{Malbet:2022lll}, assuming 100 measurements over 10 years with 1 $ \mu $as for $ 10 ^{ 8} $ sources. 

The projected sensitivities are shown in Fig.~\ref{fig:constraints}, fixing $ k / m \sim 10 ^{ - 3} $ for concreteness. While this result is unlikely to hold for such light relics, our results can easily be rescaled to any velocity. A remarkable feature of astrometric dark matter searches is their sensitivity to ultralight relics. Indeed, at the lowest feasible dark matter mass, $ m \simeq H _0 $, we estimate that current Gaia DR3 data could discover dark matter with a density as low as $ 10 ^{ - 10} $ times the known local dark matter density. 

Importantly, our method can be readily extended to ultralight vector (or tensor) dark matter. A coherently oscillating vector field will not only generate scalar metric perturbations but also sizeable vector and tensor perturbations due to the direction of the field’s oscillation. As discussed in~\cite{Nomura:2019cvc}, the timing residual for vector dark matter exhibits a directional dependence, in contrast to scalar dark matter. Consequently, we expect a similar angular dependence in astrometric probes, which could allow us to distinguish between scalar and vector dark matter. A detailed study of this possibility is left for future work.

Astrometry is complementary to cosmological probes of ultralight dark matter, which use a combination of cosmic microwave background and large-scale structure data to probe relics in a similar mass range~\cite{Hlozek:2014lca,Hlozek:2016lzm,Hlozek:2017zzf,Lague:2021frh,Vogt:2022bwy,Rogers:2023ezo,Lague:2023wes}. An astrometric search for dark matter could dramatically change the landscape of these efforts.
\vspace{0.1in}

\vspace{0.2cm}
\noindent \textbf{Acknowledgments.} 
The authors would like to thank Lam Hui, Peter Graham, Cristina Mondino, and Wei Xue for useful discussions on the density distribution of ultralight dark matter. 

\vspace{0.2cm}
\noindent \textbf{Note Added.} 
At the final stage of this work, we became aware of Ref.~\cite{Kim:2024xcr} which also considers astrometric deflection of light by dark matter.

\bibliography{references}


\clearpage

\onecolumngrid

\newpage

\widetext
 \begin{center}
   \textbf{\large SUPPLEMENTAL MATERIAL \\[.2cm] ``Astrometric Detection of Ultralight Dark Matter''}\\[.2cm]
  \vspace{0.05in}
  {Jeff A. Dror and Sarunas Verner}
\end{center}
\setcounter{equation}{0}
\setcounter{figure}{0}
\setcounter{table}{0}
\setcounter{page}{1}
\setcounter{section}{0}
\makeatletter
\renewcommand{\thesection}{S-\Roman{section}}
\renewcommand{\theequation}{S-\arabic{equation}}
\renewcommand{\thefigure}{S-\arabic{figure}}


\section{Angular Deflection Computation}
\label{app:astrodeflection}
Here, we provide the details of the derivation that were not included in the main text. We begin by calculating the first-order correction to the photon four-momentum using the geodesic equation~(\ref{eq:geopertmom}). Their solutions are,
\begin{align}
    \label{eq:p1pertfull}
    \delta p^0(\lambda) &  = \delta p^0 (\lambda_{\rm s}) + 2\omega_0 \left[  \phi ( \lambda ) - \phi ( \lambda _{ {\rm s}} ) \right]  \, , \\ 
    \delta {\mathbf{p}} (\lambda)  & =  \delta {\mathbf{p}} ( \lambda _{ {\rm s}}) -  2 \omega_0 \left[\phi(\lambda) - \phi(\lambda_{ {\rm s}})  \right] \hat{\bf n}  \, .     \label{eq:p1}
\end{align}

To calculate the observed photon four-momentum, we must first determine the integration constants, $ \delta p ^\mu ( \lambda _{\rm s} ) $. We will extract these by imposing a set of boundary conditions, two of which relate the photon worldline to the observer and source worldlines. To impose these conditions, we need the expression for the photon worldline perturbation, found by integrating the momentum equation,
\begin{equation}
    \delta {\mathbf{x}}  (\lambda) = \delta {\mathbf{x}} (\lambda _{ {\rm s}}) +  ( \lambda - \lambda_{\rm s} )  \delta {\mathbf{p}} (\lambda _{ {\rm s}})  - 2 \omega_0 \hat{\bf n}\hspace{-0.05cm}\int_{\lambda _{\rm s}}^{\lambda} \hspace{-0.05cm}\phi(\lambda') - \phi(\lambda _{ {\rm s}}) d\lambda'  \, .
    \label{eq:x1full}
\end{equation}
We calculate the source and observer location perturbations using their geodesic equations, which we parameterize using the photon's affine parameter,
\begin{align}
    \label{eq:geodesicgen}
    \frac{1}{ \omega _0  ^2 }\frac{d^2 \delta x^{\mu}_{\rm s}}{d \lambda ^2} &  = \frac{1}{ \omega _0 }\frac{d \delta u^{\mu}_{\rm s}}{d \lambda } =  - \Gamma^{\mu}_{\nu \rho} ( \lambda ) u^{\nu}_{\rm s} u^{\rho}_{\rm s}  \, ,\\ 
    \label{eq:geodesicgen2}
    \frac{1}{ \omega _0  ^2 }\frac{d^2 \delta x^{\mu}_{\rm o}}{d \lambda ^2} &  =  \frac{1}{ \omega _0 }\frac{d \delta u^{\mu}_{\rm o}}{d \lambda } =  - \Gamma^{\mu}_{\nu \rho} ( \lambda ) u^{\nu}_{\rm o} u^{\rho}_{\rm o}  \, ,
\end{align}
where the Christoffel symbols are evaluated along the zeroth order trajectory of the source and observer. Solving the differential equation with the metric given in Eq.~(\ref{eq:pertnewt}) and dropping unobservable constants yields the spatial perturbations,
\begin{align} 
     \label{eq:velocityobserver1}
   &  \delta {\mathbf{u}} _{\rm o}(\lambda )  =    \omega _0 \int^{\lambda   }_{\lambda _{\rm ref}} \hspace{-0.2cm} \nabla \phi  (t ( \lambda ') , {\mathbf{0}}  )  d \lambda '\, , \\ 
   &  \delta {\mathbf{u}} _{\rm s} ( \lambda )  =   \omega _0  \int^{\lambda   }_{\lambda _{\rm ref}} \hspace{-0.2cm}\nabla \phi  (t ( \lambda '), {\mathbf{x}} _{\rm s}  )d \lambda '  \, .
    \label{eq:velocityobserver2} 
\end{align} 
The lower bound on the integral is taken at an unphysical reference point, $\lambda _{\rm ref}$, and we have dropped an unmeasurable integration constant. Note that the integrands are implicitly functions of $ \lambda $ from the zero-order relation, $ t ( \lambda ) =\omega _0  \lambda + t _{ {\rm o}}  $. Additionally, we evaluate the perturbations of the observer at the origin, neglecting the observer displacement due to their zeroth-order velocity. Including this displacement throughout introduces corrections suppressed by the distance to the source and we have checked that they do not meaningfully influence our final result. Integrating the geodesic equations once more yields expressions for the deflection of the observer and source positions:
\begin{align} 
 \delta {\mathbf{x}}  _{ {\rm o}} & =  \omega_0\int _{ \lambda  _{ {\rm ref}}} ^\lambda  \hspace{-0.2cm}d \lambda  ' \delta {\mathbf{u}} _{\rm o} ( \lambda' )  \,,~ \delta {\mathbf{x}}  _{ {\rm s}}  =  \omega_0\int _{ \lambda  _{ {\rm ref}}} ^\lambda  \hspace{-0.2cm}d \lambda'  \delta {\mathbf{u}} _{\rm s} ( \lambda' ) \,.
\label{eq:positionobserver}
\end{align} 
Finally, we need the expression for the zeroth component of the four-velocities. For a static observer, $ \delta u ^0 _{ {\rm o}} ( \lambda ) =  \phi  ( t ( \lambda ) , {\mathbf{0}} ) $, while for a static source, $ \delta u ^0 _{ {\rm s}} ( \lambda ) =  \phi  ( t ( \lambda ) , {\mathbf{x}} _{\rm s}  )  $. 

We are now ready to calculate the four integration constants $\delta p ^\mu  ( \lambda _{ {\rm s}} ) $ by imposing four boundary conditions.

\paragraph{Photon geodesic is null.} The null geodesic condition is $ ( \eta _{ \mu\nu } + h _{ \mu \nu } ( \lambda ) ) ( p^{\mu} + \delta p ^\mu ( \lambda ) ) (  p^{\nu} + \delta p ^\nu ( \lambda ) )  = 0$ for any value of $ \lambda $. Evaluating this constraint at $\lambda = \lambda _{\rm s} $, leads to:
\begin{equation} 
  \hat{\bf n} \cdot \delta {\mathbf{p}} (\lambda _{\rm s} ) = -\delta p^0(\lambda _{\rm s} )\,. 
\label{eq:ppar}
\end{equation} 

\paragraph{Photon frequency at emission is $ \omega _0 $.} 
The emitted frequency of the photon is given in terms of the metric, photon four-momentum, and the source velocity at $ \lambda = \lambda _{\rm s} $: 
\begin{equation}
 \omega_0  =  - ( \eta _{ \mu \nu } + h _{ \mu \nu }(\lambda_{\rm s}) )  ( p^{\mu} + \delta p ^\mu (\lambda_{\rm s}) ) (   u_{\rm s}^{\nu} + \delta u_{\rm s}^{\nu}(\lambda_{\rm s}) )\,.
 \label{eq:p1lambdas}
\end{equation} 
This simplifies to:
\begin{equation} 
\delta p^0 (\lambda_{\rm s}) = \omega_0 \left[ \phi (\lambda_{\rm s}) - \hat{\bf n} \cdot \delta {\mathbf{u}}  _{ {\rm s}}  ( \lambda _{\rm s} ) \right]  \,. \label{eq:deltap0}
\end{equation}

\paragraph{Photon and observer worldlines intersect.} The perturbed photon path must intersect with the observer, which generically pushes the observed affine parameter away from $0$. However, we can set $\lambda$ at observation to $0$ even at first order using the reparametrization symmetry of the affine parameter, $\lambda \rightarrow a\lambda +b$. Then, at first order, requiring the photon intersect the observer sets $ \delta {\mathbf{x}} (0) = \delta {\mathbf{x}} _{\rm o} ( 0 )   $. Equivalently, using Eq.~\eqref{eq:x1full},
\begin{equation} 
\delta {\mathbf{x}} (\lambda _{ {\rm s}})  =     \delta {\mathbf{x}}  _{\rm o} (0) +  \lambda_{\rm s} \delta {\mathbf{p}} (\lambda _{ {\rm s}}) + 2 \omega_0  \hat{\bf n}\hspace{-0.05cm}\int_{\lambda _{\rm s}}^{0} \hspace{-0.05cm}\phi(\lambda') - \phi(\lambda _{ {\rm s}}) d\lambda'   \, . \label{eq:deltaxs}
\end{equation} 
This fixes $ \delta {\mathbf{x}} ( \lambda _{ {\rm s}} ) $; a quantity we require to apply the final boundary condition.

\paragraph{Photon and source worldlines intersect.} The perturbed photon path must intersect with the source. This requires a correction to the affine parameter at emission $ \delta \lambda _{ {\rm s}} $, which cannot be redefined away. Matching the source position to the photon location implies, 
\begin{equation} 
{\mathbf{x}} _{\rm s} ( \lambda _{\rm s} + \delta \lambda _{\rm s} ) = {\mathbf{x}} ( \lambda_{ {\rm s}}  + \delta \lambda _{ {\rm s}}) + \delta {\mathbf{x}} ( \lambda _{ {\rm s}} ) \,.
\end{equation} 
Equating the first-order correction terms, 
\begin{equation} 
\delta {\mathbf{x}} _{\rm s} ( \lambda _{\rm s} ) = - \delta \lambda _{\rm s} \omega _0  \hat{\bf n} + \delta {\mathbf{x}} ( \lambda _{ {\rm s}} ) \,, \label{eq:deltaxs1}
\end{equation} 
where $ \delta {\mathbf{x}} ( \lambda _{\rm s} ) $ is fixed from Eq.~\eqref{eq:deltaxs}. Since we already have the component of $ \delta {\mathbf{p}} ( \lambda _{\rm s} ) $ parallel to $ \hat{\bf n} $ from Eqs.~\eqref{eq:ppar} and \eqref{eq:deltap0}, we just need the perpendicular component. To this end, we take the cross product of Eq.~\eqref{eq:deltaxs1} with $ \hat{\bf n} $:
\begin{equation} 
\hat{\bf n} \times  \delta {\mathbf{p}} (\lambda _{ {\rm s}}) =   \frac{1}{ \lambda_{\rm s}}\hat{\bf n} \times \left[ \delta {\mathbf{x}} _{\rm s} ( \lambda _{\rm s} ) - \delta {\mathbf{x}}  _{\rm o} (0) \right] \,. \label{eq:deltaxs2}
\end{equation} 
Combining the results for the parallel and perpendicular components, the spatial integration constants are:
\begin{align} 
´ \delta {\mathbf{p}} ( \lambda _{\rm s} ) & = - \omega_0 \hat{\bf n} \left[ \phi (\lambda_{\rm s}) - \hat{\bf n} \cdot \delta {\mathbf{u}}  _{ {\rm s}}  ( \lambda _{\rm s} ) \right] -  \frac{1}{ \lambda_{\rm s}}\hat{\bf n} \times \left[ \hat{\bf n} \times \left[ \delta {\mathbf{x}} _{\rm s} ( \lambda _{\rm s} ) - \delta {\mathbf{x}}  _{\rm o} (0) \right] \right] \,.\label{eq:deltapi_}
\end{align} 
Eqs.~\eqref{eq:deltap0} and \eqref{eq:deltapi_} can now be input into Eq.~\eqref{eq:p1} to give the photon four-momentum at $ \lambda = 0 $ in the coordinate frame.

Solving the parallel transport equation~(\ref{eq:geodesicfore}), we find
\begin{align} 
\delta e_{\hat{a}}^i & =   \phi (t, {\mathbf{0}} ) \delta_{\hat{a}}^i \,,\\ 
\delta e ^0 _{ \hat{ a}}   & = \omega _0 \int _{ \lambda _{\rm ref}} ^\lambda   \hspace{-0.1cm} \partial _{ \hat{ a }} \phi   ( t' ( \lambda )  ,{\mathbf{0}} ) d \lambda ' \,,
\end{align} 
up to omitted integration constants which are time-independent and unobservable. 

We now have the explicit expressions for $\delta p^{\mu}(0)$ and $\delta e_{\hat{\alpha}}^{\mu}(0)$. Continuing with Eq.~(\ref{eq:deltan2}), it can readily be shown that the first three terms in the square brackets cancel when taking the limits $x_{\rm s} \ll m^{-1}$ and $x_{\rm s} \gg m^{-1}$ because they do not depend on the distance to the source, $x_{\rm s}$. Using Eq.~(\ref{eq:deltan2}) from the main text together with Eqs.~(\ref{eq:velocityobserver1}) - (\ref{eq:positionobserver}), one can show that the remaining term is given by:
\begin{align}
    -\frac{\delta p^j(0) \delta^i_{\hat a}}{\omega_0 } \hat{\theta}^{\hat a} &= \; \frac{(\delta {\bf x}_0(0) - \delta {\bf x}_{\rm s}(\lambda_{\rm s}))}{x_{\rm s}} \cdot \hat{\theta}\,, \\
    & \simeq \; \frac{1}{2 x_{\rm s} m }\phi_{\rm c} \left[ \hat {k} (0) \cdot {\hat \theta}\sin(2 m t_0 + \alpha(0)) - \hat {k} (x_{\rm s}) \cdot {\hat \theta}\sin(2 m (t_0 -  x_{\rm s}) + \alpha(x_{\rm s})) \right] \, ,
\end{align}
When $ x_{\rm s} \gg m^{-1}$, this expression can be neglected since it is suppressed by the denominator contribution. When $x_{\rm s} \ll m^{-1}$, the phases and dark matter field momenta between the emission and detection match, $\alpha (0) \simeq \alpha(x_{\rm s}) $ and $\hat{k}(0)=\hat{k}(x_{\rm s})$. We may expand the expression using a Taylor expansion: 
\begin{equation}
    - \frac{\delta p^j(0) \delta^i_{\hat a}}{\omega_0 } \hat{\theta}^{\hat a}  \; \simeq \; \frac{k(0)}{m} \phi_{\rm c} \cos(2 m t + \alpha(0)) \, ,
\end{equation}
where we replaced the observation time with $t_0 \rightarrow t$ and neglected a small contribution suppressed by $m^2$. Importantly, this leads to the difference in deflections given by Eq.~(\ref{eq:deflectiondifference}). We note by taking the difference of the $m x_{\rm s} \ll 1 $ and $m x_{\rm s} \gg 1 $ limits, we arrived at a final result insensitive to the distance to the source $x_{\rm s}$.

\section{Sensitivity}
\label{app:sensitivity}
In Eq.~\eqref{eq:qth} we presented an expression for the sensitivity of a regular-cadence search to a secular proper motion and oscillations in the angular positions of quasars. In this section, we derive these expressions using the log-likelihood ratio test. Consider a single quasar whose position is measured with a Gaussian instrumental sensitivity $ \sigma _\theta $, every $ \Delta t $, for a total time span of $ T $. We model the expected signal as a deterministic function, $ \bar{\theta} ( t ) $, equal to $ \omega _\theta t $ for a search for a secular proper motion. The likelihood of $ N _t $ measurements is,
\begin{equation} 
{\cal L} ( \bar{\theta}    | \left\{ \theta _j \right\} ) = \frac{1}{ ( 2 \pi \sigma _\theta  ^2 ) ^{  N _t / 2}} \exp \left[ - \frac{1}{ 2 \sigma _\theta  ^2 } \sum _{j = 1 } ^{N _t } ( \theta  _j -  \bar{\theta} ( t _j )    )  ^2 \right]\,.
\end{equation} 
The log-likelihood ratio between our signal model and background hypothesis ($ \bar{\theta}   = 0 $) is,
\begin{equation} 
q \equiv \log \frac{ {\cal L} ( 0 | \left\{ \theta _j \right\} ) }{ {\cal L} ( \bar{\theta}    | \left\{ \theta _j \right\} ) } = \frac{1}{ 2 \sigma _\theta  ^2 } \left[ \sum _{j = 1 } ^{N _t } ( \theta  _j -  \bar{\theta} ( t _j )  )  ^2   - \theta  _j ^2 \right] \,.
\end{equation} 
To determine the projected sensitivity, we employ the Asimov dataset~\cite{Cowan:2010js}, where the observed dataset is assumed to be given by the background-only hypothesis, $ \theta  _j  = 0 $. In this case, the sums can be done analytically for both models. Assuming $ T / \Delta t \gg 1 $,
\begin{equation} 
q \simeq \frac{1}{6 \sigma _\theta  ^2} \frac{\omega _\theta  ^2   }{ 6  }  \frac{ T ^3 }{ \Delta t   } \, .
\end{equation} 
We obtain our projected 95\% confidence level limit by setting $ q = q _{ {\rm th}} \simeq  2.7$. For $ N _q $ quasars, the experimental sensitivity grows as $ 1 / \sqrt{ N _q } $. This yields Eq.~\eqref{eq:qth} in the main text.

\end{document}